\documentclass[11pt]{article}
\usepackage{jcappub}
\usepackage{graphicx}
\usepackage{caption}
\usepackage[utf8x]{inputenc}
\usepackage{color}
\usepackage{subfig}
\usepackage{multirow}
\usepackage{slashed}
\usepackage{subfig}
\usepackage{bmpsize}
\usepackage{epstopdf}

\title{What Could be the Observational Signature of Dark Matter in Globular Clusters?}

\author[a,b]{Elaine C. F. S. Fortes}%
\affiliation[a]{Universidade Federal do Pampa\\
	Rua Luiz Joaquim de Sá Brito, s/n, Promorar,\\ Itaqui - RS, 97650-000,
	Brazil}
\affiliation[b]{Instituto Nacional de Pesquisas Espaciais\\
	Av. dos Astronautas, 1758 - Jardim da Granja, \\São Jose dos Campos, SP, 01506-000, Brazil}

\author[b]{Oswaldo D. Miranda}%

\author[c]{Floyd W. Stecker}%
\affiliation[c]{Astrophysics Science Division, \\ NASA Goddard Space Flight Center, Greenbelt, MD 20771
}

\author[b]{Carlos A. Wuensche}%

\emailAdd{elainefortes@unipampa.edu.br}
\emailAdd{oswaldo.miranda@inpe.br}
\emailAdd{floyd.w.stecker@nasa.gov}
\emailAdd{ca.wuensche@inpe.br}

\abstract{Here we investigate the possibility that  some globular clusters (GCs) harbor intermediate mass black holes (BH) in their centers and are also embedded in a low-mass dark matter (DM) halo.
	
	Up to date, there is no evidence on whether or not GCs have DM in their constitution. For  standard cold DM cosmology, it is expected that GCs form with their own DM halos. Other studies  investigate the possibility that GCs were  initially  embedded in massive DM halos that evolved during the cluster lifetime.
	
	An additional intriguing question  is  related to the existence of intermediate mass black holes (IMBH) in the  of GCs.  The determination of whether GCs hold IMBHs would be able to answer important questions about GCs formation and the circumstances that gave rise to the IMBHs. DM \& IMBH in the context of GCs are interesting subjects to be studied and we will perform such studies here, assuming the coexistence of both of them in some GCs.
	
	Having such information, we perform the study possible DM signals from GCs. One important subject to be studied is the DM density profile.  In the inner regions of GCs, the DM density profile is still an open question of fundamental importance to DM studies, specially for the study of radio and $\gamma$-ray emission from DM annihilation in such regions. Here we consider the case of IMBHs in the inner regions of three GCs: 47 Tuc, NGC 6266 and $\omega$ Cen. The existence of a black hole (BH) directly affects the matter distribution in its vicinity. These effects are significant for IMBHs as well as supermassive black holes (SMBHs) and can lead to large DM overdensities called spikes. In this paper, we direct our studies to the radio synchrotron emission from DM annihilation with spike profiles around possible IMBHs in GCs.  We present our results for synchrotron emission produced by DM annihilation via $b\bar{b}$ channel. We use the best-fit DM mass of 34 GeV and annihilation channel in $b\bar{b}$ used to explain the Galactic center ``excess" and perform our analysis for the estimated radio flux. We direct our attention specially to the $\gamma$-ray emission from 47 Tuc. When considering a scenario with includes a BH and the combined analysis of  multi-wavelength studies, we conclude that some parameter choice as the annihilation cross section $\langle\sigma v\rangle$ used to fit the $\gamma$-ray excess differs by many orders of magnitude from the one necessary to fit the radio observations for this globular cluster if we consider the BH scenario. }

	
\begin{document}	
\maketitle

\section{Introduction}
\label{sec:intro}

${}$

Despite great advances in Particle Physics, Astrophysics and Cosmology, we still have no knowledge about the nature or composition of dark matter (DM). For particle DM, we have little knowledge of range of parameter space of the model to which it might belong. Together with these uncertainties, there exist questions related to  DM density profiles. The structure of DM halos are the subject of ongoing numerical and analytical studies.

N-body simulations have been used to study halo formation. These studies have shown that, although the spherically-averaged density profiles of the simulated halos are very similar, their profiles are significantly different from the single power laws predicted by theory. This conclusion holds even considering different halo masses or different cosmological models \cite{DelPopolo:2009xj}.  The N-body profiles are characterized by a decline obeying an $r^{-3}$ proportionality at large radii and a cusp profile of the form $r^{-\gamma}$, the so called $\gamma$-models \cite{Widrow:2000dm1,Widrow:2000dm, Graham:2005xx}.

Analytical studies of DM halos predict that the density profile of a virialized halo should obey the power-law $\rho\propto r^{-9/4}$. The $r^{-9/4}$ refers to an approach done  semi-analytical models, which included a variety of physical ingredients to describe the collapse and virialization of halos that are spherically symmetric, have suffered no major mergers and have undergone quiescent accretion.   A great part of the analytical work is based on the secondary infall model (SIM) introduced by Gunn \& Gott (1972),  Gott (1975)   and Gunn (1977) \cite{Gunn1, Gott, Gunn2}.

 Similar results were also obtained when studying density profiles around density peaks and considering an isotropic velocity dispersion in the core of collisionless halos \cite{DelPopolo:2009xj}.

The most studied predictions for cold DM establish that DM halos should exhibit steep central cusps, with their density proportional to $r^{-\gamma}$ . Semi-analytical approaches indicate the existence  of a cusp having $1 < \gamma< 2$ \cite{Navarro1, Navarro2, Zhao, Cole:1995ep}. However, other simulations had find values of  $\gamma$ ranging from 0.3, 1 and 1.5 \cite{Syer:1996rd, Hoffman:1985pu, Kravtsov:1997dp, Navarro:1995iw}.

There are five common numerical profiles very well studied in the literature, namely Navarro-Frenk-White (NFW), Einasto (EIN), Isothermal (ISO), Burkert (BUR) and Moore (MOO) \cite{Navarro:1995iw, Graham:2005xx, Begeman:1991iy, Bahcall:1980fb, Burkert:1995yz}. All of them find a  cusp, assuming spherical symmetry. Their  forms as functions of the galactocentric coordinate \textbf{r} can be found at \cite{Buch:2015iya}.  An environment that could enhance and change such distribution profiles is the surroundings of compact objects such as black holes (BH).

  The presence of intermediate mass black holes (IMBH) and supermassive black holes (SMBH) modifies the distribution of dark matter in its vicinity, making  such regions, very promising locations to detect DM indirectly due to the constant evolution of BH, such as growth,
 mass accretion, etc.  DM annihilation or decay products could be searched for in the vicinities of BH since these massive objects could induce overdensities, called ``spikes" or lighter overdensities called ``mini-spikes". Indirect searches  could also utilize the detection of DM via the detection of  fluxes of secondary particles such as $\gamma$-rays, electrons, positrons and neutrinos in such regions.

In the literature there are many theoretical studies of  the distribution of DM in the surroundings of a BH. Gondolo and Silk first studied DM spike density profile in the Galactic center, where a supermassive black hole $(2-4)\times 10^{6} M_{\odot}$ dominates the gravitational potential \cite{Gondolo:1999gy, Gondolo:1999ef, Gondolo:2000pn}.

One important recent analysis invoking the ``spiky'' nature of DM where performed at Ref. \cite{Fields:2014pia}, where it was shown the huge enhancement to the $\gamma$-ray flux from the Galactic center assuming such spike and considering that the $1-3$~GeV excess observed in Fermi data could be due to DM annihilations. More recent analysis trying to prove the existence of a DM density spike allied to the radial dependence at Galactic center where performed by Lacroix et al. \cite{Lacroix:2013qka}, who also presented a solution for the cosmic ray diffusion on very small scales, assuming a radial dependence based on NFW $+$ spike profile.

Following the studies of Gondolo, in 2000s, Silk and Zhao analysed possible lighter overdensities  around IMBHs remnants of Pop III stars and presented the first estimative of mini-spikes for the IMBH in the Milky Way \cite{Zhao:2005zr}. A more recent study used the idea of these lighter overdensities due to DM annihilation in the surroundings of IMBH  to account for the $\gamma$-ray excess in the Galactic center \cite{Lacroix:2017uqn}.

In fact, the DM density profile around IMBHs is still unknown, despite many suggestions of strong enhancements for it. On one hand, some authors claim that effects of dynamical relaxation by stellar interactions, merger between two halos, and the loss of DM falling into the IMBH  may smooth the already produced spike \cite{Vasiliev:2008uz, Ullio:2001fb}. On the other hand, all the range of dynamical effects of IMBH-DM interactions, such as black hole growth time scale, core relaxation, time from stellar dynamical heating and adiabatic response of DM has not been fully explored.

In this paper we will revisit the already studied spike profile used to explain the $\gamma$-ray emission from the globular cluster 47 Tuc \cite{Brown:2018pwq}.

We perform our simulations with this profile considering  the synchrotron emission  from DM annihilations in three globular clusters, namely, 47 Tuc, $\omega$ Cen and NGC 6266.
We will study 34 GeV DM particles which annihilates in $b\bar{b}$ channel, motivated by best-fit of the $\gamma-$ray ``excess"  for the Galactic center (See, e.g., Ref.~\cite{Brown:2018pwq})\footnote{Although the millisecond pulsar hypothesis for the excess $\gamma$-rays from the Galactic center has been favored, the DM hypothesis has recently been revived~\cite{le19}.}. We  then compare our results with the existing constraints from radio signatures for globular clusters considering the scenarios with and without BHs. We also compare the parameter results with the ones used to explain the $\gamma$-ray emission from the globular cluster 47 Tuc considering the possible existence of an IMBH \cite{Brown:2018pwq}.

The paper is organized as follows:  in section \ref{sec:Globular} we present the globular clusters to be studied, its characteristics and relevance for our analysis. In section \ref{sec:Profiles} we illustrate the DM density spike profile and its parameters used in our simulations. In section \ref{sec:Synchrotron}, we review the techniques for calculation of synchrotron emission from DM annihilations.  In section \ref{sec:Numerical} we present our results of the synchrotron flux for the chosen globular clusters and in the final section \ref{sec:Conclusion} we summarize and highlight our conclusions.

\section{Globular Clusters}
\label{sec:Globular}
${}$

Globular clusters (GCs)  consist of dense spherical concentrations of stars, with typical half-light radii of 1-10 pc and belong to the list of the oldest astrophysical objects with typical ages of 11.5 Gyr to 12.5 Gyr. The conditions under which they are formed and their various possible formation mechanisms, both inside and outside of galaxies, are open questions. These questions involve the various formation, assembly, and evolution processes of the larger galaxies \cite{Bastian}.

When we talk about DM in GCs, up to date, there is no evidence on whether or not GCs are embedded in a DM halo.  In 1968,  Peebles \& Dicke first proposed a primordial scenario of GCs   within individual DM halos \cite{Peebles:1968nf}. But, until today, the existence of DM halo in GC formation has been debated \cite{Heggie:1995bk, Ibata:2012eq}. The existence of DM in GCs could be  inferred from the kinematics of GCs stars as well. For  standard cold DM cosmology, it is expected that GCs form with their own DM halos. Other scenarios include the possibility that GCs were  initially  embedded in massive DM halos that evolved during the cluster lifetime.

Another polemic question involving GCs is related to the IMBH possibly host by them. One of the natural places to produce IMBH would be the centers of GCs \cite{Miller:2001ez}. During the past years, several methods have been taken into consideration for proving the existence of IMBH.
The main efforts trying to explore the presence of IMBHs in GCs, constraining their mass  utilize two
main approaches which investigate dynamical (kinematic measurements such as radial velocities
and proper motions) and accretion signatures \cite{Anderson:2009ku}. Some attempts try to associate the ultraluminous X-ray sources with IMBH \cite{Miller:2004bu} and other ones try to find faint IMBHs in the radio band \cite{Maccarone:2004md},  but until today, there is no  conclusive evidence for their existence.

But there are also some claims of observational evidence for these IMBHs in GCs \cite{Gebhardt:2005cy}. The prediction that IMBH (compact objects which mass is in the range  $100 M_{\odot}\lesssim M_{BH}\lesssim 10^{5} M_{\odot}$) might exist in  low-mass  stellar  systems  such  as  GCs  was made in the 1970's by Wyller \cite{Wyller}, Silk \& Arons \cite{Silk} and Frank \& Rees \cite{Rees}. Silk \& Arons studied X-ray sources in a large sample of GCs concluding that the observed flux of X-rays could be explained considering a mass accretion of  $10^{2}-10^{3} M_{\odot}$ of a central BH . This  discovery promoted the black-hole hunting in GCs using mainly X-ray and radio emission \cite{Lutzgendorf:2013csa}. The GCs have also been studied several times across the electromagnetic spectrum and led to the discovery of a large number of millisecond pulsars (MSPs).

The existence of a central BH in a globular could be inferred from several properties of the cluster, such as the total mass, total luminosity and velocity dispersion profile. Using these properties, it is possible to establish correlations  such as the well known $M-\sigma_{*}$, $M_{*}-M_{tot}$, $M_{*}- L_{tot}$, which are satisfied by galaxies and which can also be satisfied by IMBH, where some of them ($M- \sigma_{*}$) are stronger than others \cite{Lutzgendorf:2013csa}.

The  $M- \sigma_{*}$ relation can be obtained from theoretical N-body simulations in combination with dynamical methods \cite{Lu:2011mya}. Recent radio continuum observations of some GCs have put an upper limit on central IMBHs. However, we emphasize that the existence of IMBH in GCs is still a subject of debate by many groups and it's  a substantially  controversial topic.

Stronger evidence concerning the presence of these massive objects in GCs may be unveiled by future experiments such as LISA \cite{Lisa}, which will be the first dedicated space-based gravitational wave detector. More theoretical work will also be very useful in investigating the existence of central IMBH in GCs.

Ref. \cite{Gribel} explored a model which unified the cosmic star formation rate with the local galactic star formation rate via a hierarchical structure formation scenario. As a consequence of the model, it was predicted that dark matter halos of large mass could contain a number of halos of much smaller mass, being able to form structures very similar to GCs. In particular, \cite{sollima} estimate the fraction and distribution of dark matter in the innermost regions of NGC 6218 (M12) and NGC 288. They concluded that there is a large mass fraction of non-luminous matter in these objects.

In this article we concentrate our efforts studying  three GCs, namely, 47 Tuc also known as NGC 104, $\omega$ Cen, also known as NGC 5139 and M62, also known as NGC 6266. 47 Tuc and $\omega$ Cen are the two most concentrated and massive GCs in our galaxy. The observations of Fermi-LAT satellite very early discovered that 47 Tuc and $\omega$ Cen were $\gamma$-ray bright \cite{collaboration:2010bb}~\footnote{On the other hand, some atypical properties of  $\omega$ Cen suggest that it might be the remnant core of a dSph. If this statement is confirmed, $\omega$ Cen  is considered to be the best place to search for DM annihilation  due to its proximity to Earth and since it   contains  high density of DM as compact dwarf galaxies and also emits $\gamma$-rays \cite{Brown:2019whs}.}. NGC 6266 is among the ten most massive and luminous GCs in the galaxy. Simulations suggest that these clusters may have an IMBH on their core.

 These three GCs were observed by the Hubble Space Telescope (HST). The results of these observations placed limits on the mass a possible IMBH \cite{Lu:2011mya}.  X-ray and radio observations were  used to settle limits on the mass of a central accreting black hole for 47 Tuc. \cite{Grindlay:2001xh}.

NGC 6266 was found to be the most suitable object to search by HST which  measured this cluster's internal proper motion dispersion profile and them compared the results to those ones produced by N-body simulations of NGC 6266 with/without an IMBH \cite{McNamara}.

${}$

In Table \ref{Tab1}, we present the main characteristics for these GCs.

${}$
\begin{table}[h]
	\centering
	\begin{tabular}{ |p{4.5cm}||p{2.3cm}|p{2.3cm}|p{2.3cm}|  }
		\hline
		\multicolumn{4}{|c|}{Globular Clusters} \\
		\hline
		& 47 Tuc &$\omega$ Cen &NGC 6266\\
		\hline
		\textit{l} (Degrees)  & 305.9    &309.1&   353.6\\\hline
		\textit{b} (Degrees) &   -44.9  & +15   &+7.3\\\hline
		\textit{d} (kpc) &4.59 & 5.21&  7.05\\\hline
		$t_{BH}$(Gyr) & 11.75  & 11.52   & 11.78\\\hline
		$M_{BH} (M_{\odot})$ & 2.3$\times 10^{3}$  & 5$\times 10^{4}$   & 2$\times 10^{3}$\\\hline
		$\sigma_{*}$ (km/s) & 10  & 10  & 15\\\hline
	\end{tabular}
	\caption{Data from 47 Tuc, $\omega$ Cen  and NGC 6266. The parameter \textit{l} denotes the Galactic latitude of the GC, \textit{b} denotes Galactic longitude, \textit{d} is the distance of the GC from Earth, taken from Refs.~\cite{Lu:2011mya,chen,mcdonald,mcnamara1,mcnamara2}. The parameter $t_{BH}$ denotes the age of the IMBH which the GC might hold, $M_{BH}$ denotes the mass of the IMBH and $\sigma_{*}$ is the stellar velocity dispersion. The numerical values for these parameters can be found at Refs \cite{Loeb, Lutzgendorf:2012sz, Noyola:2008kt}. We emphasize that the BH masses used in this table are models and are not conclusively confirmed.}
	\label{Tab1}
	
\end{table}

In the next sections, we will study the impact of the existence of  IMBH in these GCs, considering a scenario of DM annihilation in the vicinity of the IMBH. For the three GCs we will use the data of Table \ref{Tab1}. We will study the implication of the DM spike enhanced distribution in the synchrotron signal of these GCs.

\section{DM Spike Density Profile in Globular Clusters}
\label{sec:Profiles}
${}$

We will use here the DM spike profile of Ref.~\cite{Brown:2018pwq} in order to study the effect of DM on halos surrounding an IMBH in a GC. We emphasize that the full range of dynamical effects of a BH was not explored in considering the spike. Here, we just focus on a well motivated dense inner spike profile and try to evaluate its effects on the synchrotron emission of DM in the GCs.  We then compare our results with the one obtained by Brown et al. (\cite{Brown:2018pwq}) to fit the $\gamma$-ray excess.

The spike profile considered here presents a sharply peaked radial dependence, which accounts for both the presence of IMBH and dynamical processes in the GC. The spike structure of this profile was motivated by the strong evidence of DM mass with 47 Tuc. Brown et al. analysed the full 9-year Fermi-LAT data in attempt to find a possible explanation of the $\gamma$-ray excess, not only considering DM, but also millisecond pulsars (MSP). Approximations like this were already used to study a DM population clustered in the vicinities of a SMBH at the center of Centaurus A \cite{Brown:2018pwq, Brown:2016sbl}.

The existence of a spike enhances the $\gamma$-ray signal from DM annihilation. It can be represented by

\begin{equation}
\hbox{$\rho$}(r)=\left\{\begin{array}{lll}  0 & \hbox{} & r< 2R_{s} \\\dfrac{\rho_{sp}(r)\rho_{sat}}{\rho_{sp}(r) + \rho_{sat}} & \hbox{} & 2R_{s}\leqslant r < R_{sp} \\\rho_{0}\left(\dfrac{r}{R_{sp}}\right)^{-5} & \hbox{} &  r\geqslant R_{sp} \end{array}\right.
\end{equation}
where $R_{s}=2 G M_{BH}/c^{2}$ denotes the Scharzschild radius,  
$R_{sp}=G M_{BH}/\sigma_{*}^2$ \cite{Peebles} denotes the radius of the spike, 
$G$ denotes the Newton's constant of gravity, 
$M_{BH}$ is the mass of the black hole,
$\sigma_{*}$ is the stellar velocity dispersion,
$\rho_{sp}(r)=\rho_{0}(r/R_{sp})^{-3/2}$ and 
$\rho_{0}\approx (3-\gamma_{sp})M_{BH}/(4\pi R_{sp}^{3})$.
The parameter $\gamma_{sp}=(9-2\gamma)/(4-\gamma)$ is expected to be between 2.25 and 2.5, taking into account that $0 <\gamma < 2$. The parameter $\rho_{sat}$ is  the saturation density established by DM annihilations. In this case, this corresponds to make $\rho_{sat}=\rho_{sat}^{ann}$, where
\begin{equation}
\rho_{sat}^{ann}=\dfrac{m_{DM}}{\langle\sigma v\rangle t_{i}},
\end{equation}
basically establishing $\rho_{sat}=\rho_{sat}^{ann}$ corresponds to the equality of characteristic annihilation time and infall time ($ t_{i}$) of DM towards the IMBH. Here, we will assume conservatively that $t_{i}=t_{BH}$, where $t_{BH}$ denotes the age of the black hole, $m_{DM}$ is the mass of DM candidate and $\langle\sigma v\rangle$ is the annihilation cross section.

 For $r\geqslant R_{sp}$ Brown et al.~\cite{Brown:2018pwq} assumed a radial dependence $r^{-5}$ (from tidal stripping). This profile is normalized by requiring that the mass inside the spike $M_{sp}$ be of order of the BH mass, resulting that the total DM mass in the cluster is $\sim 1 \%$ of the mass of 47 Tuc.  In the end, the authors fit 47 Tuc's $\gamma$-ray spectrum considering MSP+DM spike.  Their best-fit solution had a DM candidate with mass 34 GeV and $\langle\sigma v\rangle\sim 6\times 10^{-30}$ cm$^{3}$/s.

 There are some comments and replies to comments related to MSP \& DM interpretation in Brown's papers \cite {Bartels:2018qgr,  Brown:2019teh}.  In the reply to comments they conclude  that such discussion motivates a deeper radio study of  47 Tuc. In this paper, we present a first analysis of the radio signature from dark matter annihilation, studying the same channel proposed by \cite{Brown:2018pwq} to explain the gamma excess in 47 Tuc. We will also present the theoretical radio flux for other two GCs, namely, $\omega$ Cen and NGC 6266.

  \section{Synchrotron/Radio Flux From Dark Matter Annihilation}
\label{sec:Synchrotron}
${}$

In this section we revisit the techniques to calculate the radio flux originated from the propagation of electrons/positrons resulting from DM annihilation processes. Then we compute the resulting synchrotron emission for three GCs. The radio flux emission depends mainly on the magnetic field strenght ($B$), on the DM mass ($m_{DM}$), annihilation channels and density profile, $\rho(r)$, and on the density of the ionized gas ($n$).

The integrated synchrotron flux density produced by a generic distribution reads:
\begin{equation}\label{Fluxo}
 S_{\nu}(\nu)\approx \dfrac{1}{D^{2}}\int_{0}^{R_{a}}dr r^{2}j_{\nu}(\nu, r),
\end{equation}
where $R_{a}$ is the radial extent of the region of interest \cite{Storm:2016bfw}. The parameter $D$ is the distance from the GC to Earth. The result is usually given in Janskys.
In eq. (\ref{Fluxo}), $j_{\nu}$ denotes the synchrotron emissivity, which can be expressed as
\begin{equation}\label{jnu}
 j_{\nu}(\nu,r)=2 \int_{m_{e}c^{2}}^{m_{DM}}dE \dfrac{dn_{e}}{dE}(E,r)P_{\nu}(\nu,E)
\end{equation}
where  the factor 2 takes into account the electrons and positrons, $ \dfrac{dn_{e}}{dE}(E,r)$ is the electron equilibrium spectrum and $P_{\nu}(\nu,E)$ denotes the synchrotron power for a certain frequency $\nu$. Considering the isotropy in the distribution of relativistic electrons with energy $E$ and uniform magnetic field, we develop the steps necessary to obtain $P_{\nu}(\nu,E)$. We  consider the analysis presented in Ref. \cite{Buch:2015iya}. First we have that:

\begin{equation}\label{Psyn}
 P_{\nu}(\nu,E,\alpha)=\dfrac{\sqrt{3}e^{3}B \sin \alpha F(x)}{m_{e}c^{2}},
\end{equation}
with $x=\nu/\nu^{\prime}$, $\nu^{\prime}=\nu_{c}\sin \alpha /2$ and $\nu_{c}=3 e B \gamma^{2}/(2\pi m_{e}c)$,  $e$ denotes the electric charge, $B$ is the magnetic field, $c$ is the velocity of light, $\gamma$ is the Lorentz factor, related to the energy of a single electron by $E=\gamma m_{e}c^2$ and $F(x)$ is given by~\footnote{ We note that reference \cite{Storm:2016bfw} considers a series approach for $F(x)$. In their approximation, $s=x/\alpha$ and $F(s)\approx 1.25s^{1/3}\exp(-s)[648 + s^{2}]^{1/12} $. }

\begin{equation}\label{Fx}
F(x)=x \int_{x}^{\infty} K_{5/3}(x^{\prime})dx^{\prime},
\end{equation}
where $K_{n}$  the  modified  Bessel  function  of  the  second  kind and  order $n$.  As the next step we need to average the  randomly oriented magnetic field over the pitch angle $\alpha$ and after, we can express  $P_{\nu}(\nu,E)$ as:
\begin{equation}\label{Psyn1}
 P_{\nu}(\nu,E)=\dfrac{1}{2} \int_{0}^{\pi} d\alpha \sin \alpha P\nu(\nu,E,\alpha),
\end{equation}
and finally we have
\begin{equation}\label{Psyn2}
 P_{\nu}(\nu,E)=2\sqrt{3}\dfrac{e^{3}B}{m_{e}c^{2}}y^{2}\left[K_{4/3}(y)K_{1/3}(y)-\dfrac{3}{5} y \left(K_{4/3}^{2}(y)-K_{1/3}^{2}(y) \right)\right],
\end{equation}
where $y=\nu/\nu_{c}$.
The term  $\dfrac{dn_{e}}{dE}(E,r)$ in the eq. (\ref{jnu}) is the electron equilibrium spectrum.

In an astrophysical medium, the diffusion and energy losses modify the injection spectrum from DM annihilation or decay. In the end, all of these mechanisms should be taken into account and the electron equilibrium spectrum is obtained from the diffusion equation
\begin{equation}\label{Dif}
 \dfrac{\partial}{\partial t}\dfrac{dn_{e}}{dE}= \triangledown\left[ D(E,r)\triangledown \dfrac{dn_{e}}{dE}\right]\\
  + \dfrac{\partial}{\partial E}\left[ b_{loss}(E,r)\dfrac{dn_{e}}{dE}\right]+Q(E,r),
\end{equation}
where $Q(E,r)$ is the source term, $D(E,r)$ is the coefficient for spatial diffusion and $b_{loss}$ is the loss term. If we neglect the diffusion coefficient term, we have
 \begin{equation}\label{Psyn2b}
\dfrac{dn_{e}}{dE}(E,r)=\dfrac{ \langle\sigma v\rangle \rho(\, {\rm \textbf{r}})^{2}}{2 m_{DM}^{2}b_{loss}(E,r)}\int_{E}^{m_{DM}}dE^{\prime}\dfrac{dN}{dE^{\prime}_{inj}},
\end{equation}
where $\langle\sigma v\rangle$ denotes de annihilation cross section, $\rho$(\textbf{r}) is the spatial distribution of DM, $m_{DM}$ is the mass of DM candidate, $b_{loss}$ is the energy loss term and $dN/dE^{\prime}_{inj}$ is part of the source term which relates the electron injection spectrum from DM annihilations. We took $dN/dE^{\prime}_{inj}$ from ref. \cite{Buch:2015iya}, it can be also obtained from packages presented in refs. \cite{Bringmann:2018lay, Belanger:2013oya}.  The source term is expressed by:
\begin{equation}\label{source}
Q(E,r)=\dfrac{\langle\sigma v\rangle \rho(\, {\rm \textbf{r}})^{2}}{2 m_{DM}^{2}}\dfrac{dN}{dE_{inj}},
\end{equation}
The energy loss term is given by
\begin{equation}
 b_{loss}(E,r)=b_{syn}+b_{IC}+b_{brem}+b_{coul}
\end{equation}
where the terms $b_{syn}$, $b_{IC}$, $b_{brem}$ and $b_{coul}$ denote the loss by synchrotron radiation, inverse Compton scattering, bremsstrahlung and Coulomb interactions, respectively.

Since we are working with GCs in the galaxy, in energies below $\sim$ 5 GeV, ionization and bremsstrahlung dominate. At higher energies Compton and synchrotron processes dominate
\cite{Floyd}.

The energy loss due to all these processes can be expressed by \cite{Storm:2016bfw}.

\begin{eqnarray}
 b_{loss}(E,r)& \approx  0.0254\left(\dfrac{E}{1 GeV}\right)^{2}\left(\dfrac{B(r)}{1\mu G}\right)^{2}+ 0.25\left( \dfrac{E}{GeV}\right)^{2}\\\nonumber
 &+ 1.51n(0.36+log(\gamma/n))+ 6.13n(1+log(\gamma/n)/75.0),
 \end{eqnarray}
 where this loss term has units of $10^{-16}$ GeV/s, $E=\gamma m_{e}c^{2}$, $B$ denotes the magnetic field, which we consider as 10 $\mu$G in all of our analysis. In our simulations $n$ denotes the number of free electrons which corresponds to the density of ionized gas. Here we have considered $n=1\times 10^{-3}$ cm$^{-3}$. For GCs, some authors also consider the gas density expressed as an exponential profile $n=n_{0}\exp (-\, {\rm \textbf{r}}/r_{D})$, where $n_{0}$ is the initial gas density \cite{Beck:2019imo}.

 \section{Numerical Results}
 \label{sec:Numerical}
${}$

In this section we present the numerical results for the synchrotron emission theoretically predicted by our calculations for the GCs 47 Tuc, NGC 6266 and $\omega$ Cen, assuming a DM candidate with $m_{DM}=34$ GeV annihilating into a $b\bar{b}$ channel. When solving the diffusion-loss differential equation, we assume that the cooling time scale of high energy electrons and positrons is much smaller than their diffusion scale, so that the diffusion term in Eq. (\ref{Dif}) can be neglected.

\begin{figure*}[h]
\centering
\begin{tabular}{cc}
\includegraphics[width=0.95\linewidth]{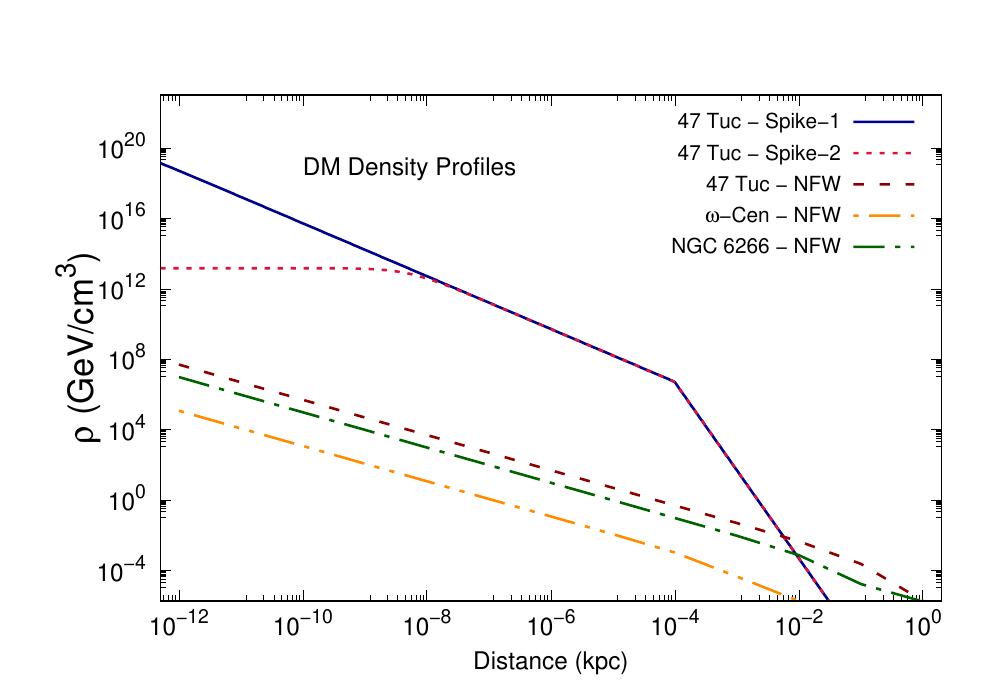}
\end{tabular}
\caption{The figure shows the spike density profile  created around the IMBH for the GC 47 Tuc. We present here two scenarios, shown in blue (Spike-1) and  dashed-pink (Spike-2). The spikes are formed within the radius of influence of the BH ($R_{sp}$) of the globular cluster 47 Tuc. We take $m_{DM}=34$ GeV, $\langle\sigma v\rangle=5\times10^{-37}$ cm$^{3}$/s for Spike-1 and $\langle\sigma v\rangle=6\times10^{-30}$ cm$^{3}$/s for Spike-2 scenario. For effects of comparison, we present in the same graph the canonical NFW DM density profile for 47 Tuc (dashed-brown), $\omega$ Cen (dashed-double dot yellow) and NGC 6266 (dashed-dot green). In this case, we consider that there are no BHs in the centers of these GCs.}
	\label{fig1}
\end{figure*}

For illustration, in Figure \ref{fig1} we show the  behavior of DM spike density profile for 47 Tuc considering two situations. Spike-1 denotes the parameters we used to fit the radio spectrum of this GC using the existing experimental limits and upper bounds on 47 Tuc and also assuming the existence of a IMBH. Spike-2 denotes a best-fit scenario of $\gamma$-rays presented by \cite{Brown:2018pwq} assuming the existence of a IMBH, where $\langle\sigma v\rangle=6\times 10^{-30}$ cm$^{3}$/s. The spikes are formed with the typical radius of influence of the BH, $R_{sp}=G M_{BH}/\sigma_{*}^2$. Out of the region of influence, the density goes as  $\rho \propto r^{-5}$.

Taking the set of parameters described in Table \ref{Tab1} together with those described in Section \ref{sec:Synchrotron}, in order to fit the radio spectrum with a DM candidate annihilating in $b\bar{b}$ with $m_{DM}=34$ GeV, we find that $\langle\sigma v\rangle$ must be $\sim 5\times 10^{-37}$ cm$^{3}$/s. This number was obtained considering the presence of an IMBH in 47 Tuc.

In the same figure we also present the canonical NFW profile for the GCs 47 Tuc, $\omega$ Cen and NGC 6266. In this case, we consider that these GCs do not have black holes in their centers.
For calculating the NFW density profile we had used the relation $\frac{\rho(r)}{\rho_{crit}}=\frac{\delta_{c}}{(r/r_{s})(1+r/r_{s})^2}$, where $r_{s}$ is the scale radius and $\delta_{c}$ is the characteristic density. The parameter $\rho_{s}$ can be expressed as $\rho_{s}=\rho_{crit}\cdot \delta_{c}$, where $\rho_{crit}=8.54\times 10^{-30}$ g/cm$^3$.

Another step in the calculation of the DM density profile is related to the virial radius and the concentration parameter $c$. This parameter is expressed as $c=r_{200}/r_{s}$, where $r_{200}$ denotes the virial radius. The characteristic density $\delta_{c}$ is given by \cite{Navarro:1996gj}

\begin{equation}\label{a4}
\delta_{c}=\frac{200}{3}\frac{c^3}{[\ln(1+c)-c/(1+c)]}
\end{equation}
One of the definitions of the parameter $c$ establishes that
$c=\log(r_{t}/r_{c})$, where $r_{c}$ denotes the core radius and $r_{t}$ denotes the tidal radius of the globular cluster \cite{Meylan:1999yb}. For 47 Tuc, $r_{c}=0.37$ pc and $r_{t}=56$ pc \cite{Ibata:2012eq, Lane:2012qm}. So that,  $c=2.17$. The calculation of $\delta_{c}$ is immediate as we know the value of $c$ and in this case, $\delta_{c}=1451.91$. It follows that  $\rho_{s}=2.21\times 10^{-4}$ GeV/cm$^3$.

As the last step, we have that the half light radius ($r_{h}$) of 47 Tuc is $9.6\pm 0.3$ pc. Considering the that $r_{h}$ is around 2\% of the virial radius $r_{200}$, it results that $r_{200}=480$ pc, so that $r_{s}=r_{200}/c \sim 0.22$ Kpc.

So the NFW DM density profile can be expressed as

\begin{equation}\label{a7}
\rho(r)=\rho_{s}\frac{r_{s}}{r}\left(1+\frac{r}{r_{s}}\right)^{-2},
\end{equation}
with $\rho_{s}=2.21\times 10^{-4}$ GeV/cm$^3$ and $r_{s}=0.22$ Kpc for 47 Tuc.

${}$

In a similar way For $\omega$ Cen,  $r_{t}=65$ pc, $r_{s}=1.63\times 10^{-3}$ kpc and $r_{c}=3.9$ pc according to references \cite{Brown:2019whs, Henleywillis:2018ufu} which results in a $\rho_{s}=7.45\times 10^{-5}$ GeV/cm$^{3}$.

${}$

For NGC 6266, $r_{t}=18$ pc, $r_{h}=2.47$ pc and $r_{c}=1.70$ according to references \cite{Mackey:2005tu, Han}. Considering the that $r_{h}$ is around 2\% of the virial radius $r_{200}$, it results that $r_{s}=7.26\times 10^{-2}$ kpc and $\rho_{s}=1.37\times 10^{-4}$ GeV/cm$^{3}$.
We had used all those information provided above to plot the graphs related to the three NFW DM density profile.

In Table \ref{Tab2}, we give the numerical values  for the three GCs. We had obtained these values with the set of parameters described in Table \ref{Tab1} and  in Section \ref{sec:Synchrotron},  considering the presence of IMBH in the GCs. In order to fit the radio spectrum  with a DM candidate  annihilating in $b\bar{b}$ with mass $m_{DM} = 34$ GeV, for $\omega$ Cen, we find that $\langle\sigma v\rangle$ must be $\sim 6\times 10^{-36}$ cm$^{3}$/s; for NGC 6266  $\langle\sigma v\rangle$ must be $\sim 8\times 10^{-38}$ cm$^{3}$/s.

${}$
	\begin{table}[h]
	\centering
\begin{tabular}{ |p{4.5cm}||p{2.3cm}|p{2.3cm}|p{2.3cm}|  }
 \hline
 \multicolumn{4}{|c|}{Globular Clusters} \\
 \hline
 & 47 Tuc &$\omega$ Cen &NGC 6266\\
 \hline
 $R_{a}$ (pc)  & 56.7    &55.1&   30.7\\\hline
 $R_{s}$ (pc) &   2.19$\times 10^{-10}$  & 4.76$\times 10^{-9}$   &1.90$\times 10^{-10}$\\\hline
 $R_{sp}$ (pc) &9.84$\times 10^{-2}$ & 2.14 &  3.79$\times 10^{-2}$\\\hline
  $\rho_{sat}$ (GeV/cm$^{3}$) & 1.83$\times 10^{20}$  & 1.55$\times 10^{19}$   & 1.14$\times 10^{21}$\\\hline
  $\langle\sigma v\rangle$ (cm$^{3}$/s) & $5\times10^{-37}$ & $6\times 10^{-36}$ & $8\times 10^{-38}$
  \\\hline
  $\rho_{s}$ (GeV/cm$^{3}$)  & $2.21\times 10^{-4}$    & $7.45\times 10^{-5}$&   $1.37\times 10^{-4}$\\\hline
   $r_{s}$ (kpc)  & 0.22    & $1.63\times 10^{-3}$ &   $7.26\times 10^{-2}$\\\hline
   \end{tabular}
    \caption{ The parameter $R_{a}$ denotes the radial extension of the globular cluster used in the integration of Eq. \ref{Fluxo}, $R_{s}$ denotes the  Schwarzschild radius, $R_{sp}$ is radius of influence of the BH, $\rho_{sat}$ denotes saturation density for the BH in the globular clusters, and $\langle\sigma v\rangle$ represents the annihilation cross section. In the scenario with IMBH, the $\langle\sigma v\rangle$ values are determined so that the radio flux predicted by the model does not exceed the observational limits inferred for each GC. We also present the parameters $\rho_{s}$ and $r_{s}$ used to fit the NFW DM density profile for the three globular clusters.}
   \label{Tab2}
   \end{table}

\begin{figure*}[ht!]
\centering
\begin{tabular}{cc}
\includegraphics[width=0.95\linewidth]{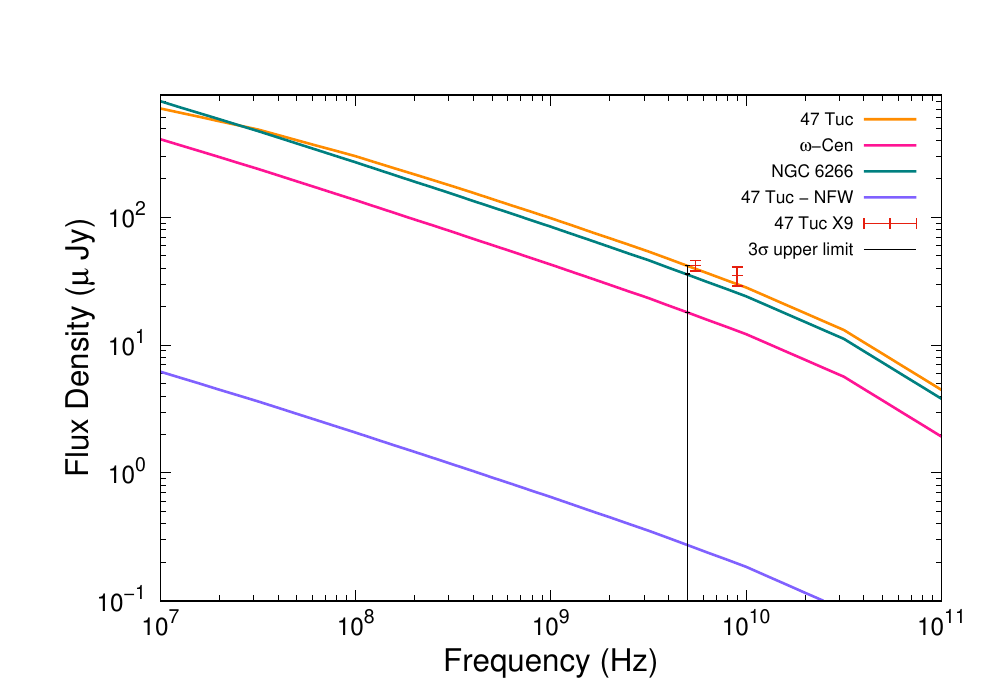}
\end{tabular}
\caption{The expected flux from DM as a function of frequency for the GCs. Our DM candidate has mass $m_{DM}=34$ GeV and annihilates in $b\bar{b}$. We had considered a scenario with an IMBH and $\langle\sigma v\rangle\sim 5\times 10^{-37}$ cm$^{3}$/s for 47 Tuc,  $\langle\sigma v\rangle\sim 6\times 10^{-36}$ cm$^{3}$/s for $\omega$ Cen and $\langle\sigma v\rangle\sim 8\times 10^{-38}$ cm$^{3}$/s for NGC 6266. The GCs have observational upper limits taken from radio continuum observations on the frequency 5 GHz (see \cite{Lu:2011mya}). The straigh line on 5 GHz indicates the 3$\sigma$-upper limits whose values are: 40 $\mu$Jy for 47 Tuc,  20 $\mu$Jy for $\omega$ Cen and 36 $\mu$Jy for NGC 6266. For comparison only, we have included the measured flux densities of 47 Tuc X9. This source is near to the center of 47 Tuc and the data are available at 5.5 GHz and 9 GHz \cite{miller-jones}. For completeness, we had included the scenario with no IMBH considering the NFW density profile for 47 Tuc.}
	\label{fig2}
\end{figure*}

In order to constrain the radio fluxes, we need to theoretically calculate their signal and estimate the sensitivity of the existing radio surveys to diffuse emission from these GCs. For the experimental values of radio flux,  we had considered the data of Ref. \cite{Lutzgendorf:2012sz, Lu:2011mya}. In Figure \ref{fig2} we give the expected flux from DM annihilation as a function of frequency for  47 Tuc, $\omega$ Cen and NGC 6266 considering the presence of an IMBH. Radio surveys of GCs have produced only upper limits for the flux density, the latest data being mainly derived by \cite{Lu:2011mya} through observations supported by the Australia Telescope Compact Array (ATCA). The upper limit flux densities at 5 GHz are presented in Fig. 2 and the values for $\langle\sigma v\rangle$ were chosen to respect these constraints.

For comparison, we present the recent measurements for the 47 Tuc X9 source \cite{miller-jones}.
This source possesses a double-peaked C IV emission line in its ultraviolet spectrum, an indicative of matter accretion. Although it was initially classified as a cataclysmic variable, the results of \cite{miller-jones} show that the spectral density is proportional to $\nu^{-0.4\pm 0.4}$ and may indicate that X9 is a black hole accreting matter. The measurements for 47 Tuc X9 at 5.5 GHz and 9 GHz are just above our estimated radio spectrum for 47 Tuc.

In examining the individual radio flux densities, we see that the dark matter annihilation cross section is much smaller than the one inferred from 47 Tuc $\gamma$-ray flux. Through the observed upper limits, $\langle\sigma v\rangle$ must be within the range $\sim 10^{-35}-10^{-37}$cm$^{3}$/s for the $b\bar{b}$ channel with $m_{DM} = 34\,{\rm GeV}$. We normalized the spike profile similarly to the one derived by \cite{Brown:2018pwq}, so that an increase in the cross section to approximate the gamma--radio signatures should be accompanied by a decrease in the amount of dark matter in the spike's influence region.

All of these results can change in order of magnitude if we  consider for instance, the  diffusion term in the diffusion-loss differential equation. In this case, the radio flux can be reduced and the soft cutoff of the signal decreases faster for lower frequencies when compared to the situation where diffusion term was neglected. Different magnetic field configurations, gas densities and annihilation cross sections, as well as the dark matter mass can also change our results. However, even exploring the parameter space defined by these variables, it is not possible to obtain signatures that are consistent with the observational radio limits when we use, e.g., $\langle\sigma v\rangle \sim 10^{-30}$ cm$^{3}$/s for 47 Tuc when we consider the presence of an IMBH.

In Figure \ref{fig2} we also present the expected flux from DM annihilation as a function of frequency for the  47 Tuc considering a scenario with no BH and a NFW DM density profile. For that flux we had considered the thermal annihilation cross section $\langle\sigma v\rangle \sim 3\times10^{-26}$ cm$^{3}$/s. For $\omega$ Cen and NGC 6266, the radio flux are smaller that the ones produced by 47 Tuc. To allow a better visualization of the figure, we didn't include  the fluxes of $\omega$ Cen and NGC 6266 due to the NFW density profile in the graph.

Considering the GC $\omega$ Cen, another limit can be obtained from reference \cite{Brown:2019whs}. On it, they presented a result obtained from $\gamma$-ray emission for this GC, concluding that if DM annihilation is responsible for $\omega$ Cen’s $\gamma$-ray emission, they could infer DM particle properties from the $\gamma$-ray spectral energy distribution. In this case, the observed spectrum is very well fitted considering a model of DM annihilation in $b\bar{b}$ with prompt photon emission and the best-fit mass is $m_{DM}= 31\pm 4$ GeV for $\langle\sigma v\rangle\sim 10^{-28^{+0.6}_{-1.2}}$  GeV/cm$^{3}$ with 68\% confidence limits, assuming a NFW distribution of DM plus a black hole.
\newpage
\section{Final Remarks}
\label{sec:Conclusion}

In this work, we present the results obtained for the radio flux of three GCs, namely, 47 Tuc, $\omega$ Cen and NGC 6266 and  the constraints for these signals obtained from radio observations considering the presence or not of an IMBH. First, to study the fluxes we had considered an enhanced DM spike density profile due to the presence of an IMBH, modeled for explaining the observed $\gamma$-ray flux of 47 Tuc \cite{Brown:2018pwq} and further extended our analyses for other two GCs. Considering that the GC holds an IMBH, we find that the annihilation cross section $\langle\sigma v\rangle$ used to explain the observed $\gamma$-ray spectrum of 47 Tuc is too high to agree with the continuum radio observations of 47 Tuc. For this GC, our simulations indicate that $\langle\sigma v\rangle\sim 5\times 10^{-37}$ cm$^{3}$/s would be in agreement with the radio surveys. We have here assumed that the DM population is dominant over the MSP population in producing the $\gamma$-ray and radio emission. A viable particle physics model is presented at \cite{Okada:2019sbb}, in which $\langle\sigma v\rangle$ is very suppressed (being much smaller than the required thermal annihilation cross section), and that could be used for DM interpretation of the$\gamma$-ray excess in 47 Tuc. 

In considering a spike profile motivated to fit $\gamma$-ray flux of 47 Tuc, we have assumed the popular scenario of a 34 GeV DM candidate annihilating into a $b\bar{b}$ channel. We have also investigated the possibility of probing DM in the inner part of  two other GCs containing IMBHs, exploring the ranges of most important parameters used in determining synchrotron flux.

Although we have performed a simplified analysis, neglecting the diffusion terms in our calculations and considering a fixed magnetic configuration, we have shown that the spike  profile studied here can leave strong signatures for the synchrotron flux from GCs.

For completeness, we had also presented the predicted radio fluxes for 47 Tuc,  $\omega$ Cen and NGC 6266 with no IMBH considering NFW DM density profile. The fluxes of  $\omega$ Cen and NGC 6266   are below the ones presented by 47 Tuc. In the calculation of all of the flux due to NFW density profile, we had used the thermal annihilation cross sections. 

In the future it would be interesting to investigate other channels for dark matter annihilation or decay, exploring different values for $m_{DM}$ and  simultaneously, taking into account the inferred $\gamma$-ray and radio constraints.

${}$

\section*{Acknowledgments}
E. C. F. S. Fortes would like to thank Marco
Cirelli and Javier Montaño Domínguez for very useful discussion. E. C. F. S. Fortes and O. D. Miranda thanks FAPESP for financial support under contract 2018/21532-4.

\end{document}